\documentclass[conference,compsocconf]{IEEEtran}
\pdfoutput=1
\IEEEoverridecommandlockouts

\usepackage{xcolor}
\usepackage[pdftex]{graphicx}
\graphicspath{{./Figures/}}
\DeclareGraphicsExtensions{.pdf}
\usepackage{wrapfig}

\usepackage{latexsym}
\usepackage{amssymb}
\usepackage{amsmath}
\usepackage[mathscr]{eucal}

\usepackage{pdfpages}
\usepackage[absolute]{textpos}
\usepackage{url}
\usepackage[colorlinks=true,urlcolor=black,linkcolor=red,citecolor=blue,hyperfootnotes]{hyperref}
\usepackage{floatrow}
\usepackage{array}
\usepackage{cite}
\usepackage{multirow}
\usepackage{ifthen}
\usepackage{expdlist}

\usepackage{algorithmicx}
\usepackage{algorithm}
\usepackage{algpseudocode}

\DeclareMathAlphabet{\mts}{U}{rsfs}{m}{n}

\DeclareMathOperator{\crd}{\mathsf{crd}}
\DeclareMathOperator{\sat}{\mathsf{sat}}
\DeclareMathOperator{\ans}{\mathsf{ans}}

\DeclareMathOperator{\ids}{\mathrm{ids}}
\DeclareMathOperator{\ip}{\mathsf{ip}}
\DeclareMathOperator{\tr}{\mathsf{tr}}
\DeclareMathOperator{\ev}{\mathsf{ev}}

\DeclareMathOperator{\cidn}{\mathrm{cidn}}
\DeclareMathOperator{\Stake}{\mathsf{Stake}}
\DeclareMathOperator{\Nonce}{\mathsf{Nonce}}
\DeclareMathOperator{\GenTime}{\mathsf{Stamp}}
\DeclareMathOperator{\PrvHash}{\mathsf{Hash}_\mathrm{old}}

\DeclareMathOperator{\id}{\mathsf{ID}}
\DeclareMathOperator{\hdr}{\mathsf{Hdr}}
\DeclareMathOperator{\True}{\mathsf{True}}
\DeclareMathOperator{\False}{\mathsf{False}}

\DeclareMathOperator{\fork}{\mathsf{fork}}
\DeclareMathOperator{\sig}{\mathsf{Sig}}

\DeclareMathOperator{\Time}{\mathsf{Time}}
\DeclareMathOperator{\Tx}{\mathsf{Tx}}
\DeclareMathOperator{\Unsure}{\mathsf{Unsure}}

\DeclareMathOperator{\ctr}{\mathsf{ctr}}

\newcommand{\zbx}[1]{\raisebox{0pt}[0pt][0pt]{#1}}

\definecolor{gr}{rgb}{0.50,0.50,0.50}
\definecolor{wh}{rgb}{1.00,1.00,1.00}

\algnewcommand{\Initial}              {\item[{\bf initialization:}]}
\algnewcommand{\Initiax}{\item[\textcolor{wh}{\bf initialization:}]}
\algrenewcommand{\algorithmicrequire}       {{\bf input:}}
\algnewcommand{\Requirx}{\item[\textcolor{wh}{\bf input:}]}
\algrenewcommand{\algorithmicensure}        {{\bf output:}}
\algnewcommand{\Outputx}{\item[\textcolor{wh}{\bf output:}]}

\algrenewcommand{\algorithmiccomment}[1]{\hfill\textcolor{gr}{$\triangleright$\ \zbx{#1}}}

\algrenewtext{If}[1]    {{\bf\algorithmicif\ }{#1}{\bf\ \algorithmicthen}}
\algrenewtext{ElsIf}[1] {{\bf\algorithmicelse\ \algorithmicif\ }{#1}{\bf\ \algorithmicthen}}
\algrenewtext{Else}     {{\bf\algorithmicelse}}
\algrenewtext{EndIf}    {{\bf\algorithmicend}}

\algrenewtext{For}[1]   {{\bf\algorithmicfor\ }{#1}{\bf\ \algorithmicdo}}
\algrenewtext{ForAll}[1]{{\bf\algorithmicfor\ all\ }{#1}{\bf\ \algorithmicdo}}
\algrenewtext{EndFor}   {{\bf\algorithmicend}}
\algrenewtext{While}[1] {{\bf\algorithmicwhile\ }{#1}{\bf\ \algorithmicdo}}
\algrenewtext{EndWhile} {{\bf\algorithmicend}}
\algrenewtext{Repeat}   {{\bf\algorithmicrepeat\ }}
\algrenewtext{Until}[1] {{\bf\algorithmicuntil\ }{#1}}

\algrenewcommand{\algorithmicindent}{6pt}

\begin{document}
\begin{textblock}{24}(1,0.2)
\noindent\tiny  This paper is a preprint; it has been accepted for publication in 2019 IEEE World Congress on Services (SERVICES),  8-13 July 2019,  Milan, Italy    \\
\textbf{IEEE copyright notice} \textcopyright 2019 IEEE. Personal use of this material is permitted. Permission from IEEE must be obtained for all other uses, in any current or future media, including reprinting/republishing this material for advertising or promotional purposes,\\ creating new collective works, for resale or redistribution to servers or lists, or reuse of any copyrighted component of this work in other works.
\end{textblock}
\bstctlcite{IEEEexample:BSTcontrol}
\title{On Blockchain Architectures for Trust-based Collaborative Intrusion Detection%
\thanks{%
\protect\begin{wrapfigure}[3]{l}{.9cm}%
\protect\raisebox{-12.5pt}[0pt][7pt]{~\protect\includegraphics[height=.8cm]{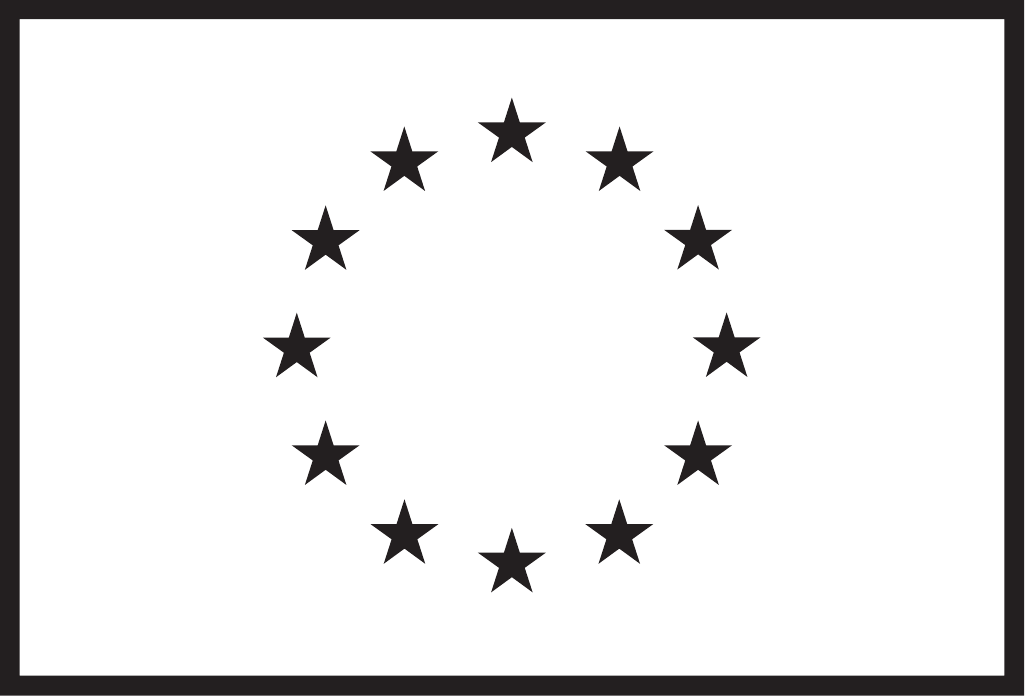}}%
\protect\end{wrapfigure}%
This project has received funding from the European Union's Horizon 2020 research and innovation programme under grant agreement no. 786698. The work reflects only the authors' view and the Agency is not responsible for any use that may be made of the information it contains.}%
}

\author{%
\IEEEauthorblockN{Nicholas Kolokotronis\IEEEauthorrefmark{1}, Sotirios Brotsis\IEEEauthorrefmark{1}, Georgios Germanos\IEEEauthorrefmark{1}, Costas Vassilakis\IEEEauthorrefmark{1} and Stavros Shiaeles\IEEEauthorrefmark{2}\vspace*{4pt}}

\IEEEauthorblockA{\IEEEauthorrefmark{1}Department of Informatics and Telecommunications, University of Peloponnese, 22131 Tripolis, Greece\\
Email: \{nkolok,\,brotsis,\,germanos,\,costas\}@uop.gr}

\IEEEauthorblockA{\IEEEauthorrefmark{2}Centre for Security, Communications and Networks Research, Plymouth University, Plymouth PL4 8AA, UK\\
Email: stavros.shiaeles@plymouth.ac.uk}
}

\maketitle

%----------------------------------------------------------%
%                      Section Change                      %
%----------------------------------------------------------%

\begin{abstract}
This paper considers the use of novel technologies for mitigating attacks that aim at compromising {\em intrusion detection systems} (IDSs). Solutions based on {\em collaborative intrusion detection networks} (CIDNs) could increase the resilience against such attacks as they allow IDS nodes to gain knowledge from each other by sharing information. However, despite the vast research in this area, trust management issues still pose significant challenges and recent works investigate whether these could be addressed by relying on blockchain and related distributed ledger technologies. Towards that direction, the paper proposes the use of a trust-based blockchain in CIDNs, referred to as {\em trust-chain}, to protect the integrity of the information shared among the CIDN peers, enhance their accountability, and secure their collaboration by thwarting insider attacks. A consensus protocol is proposed for CIDNs, which is a combination of a proof-of-stake and proof-of-work protocols, to enable collaborative IDS nodes to  maintain a reliable and tampered-resistant trust-chain.
\end{abstract}


%----------------------------------------------------------%
%                      Section Change                      %
%----------------------------------------------------------%
\section{Introduction}
\label{sec.introduction}

Cyber-security is an increasingly important aspect in the era of the {\em Internet of things} (IoT). The highly complex ecosystem of billions heterogeneous devices of weak security defenses may be exploited by attackers to launch {\em distributed denial of service} (DDoS) attacks, to steal personal data or gain full access and control to networks. Such incidents are getting more sophisticated and take place on a continuous and non-discriminatory basis. {\em Intrusion detection systems} (IDSs) constitute the basic line of defense against cyber-attacks, as they can detect suspicious behavior and deliver informative security alerts. For the recognition of large-scale and complex attacks, collaboration among the stand-alone IDSs has been developed \cite{meng15}. The term {\em collaborative intrusion detection networks} (CIDNs) refers to such network of communicating IDSs that exchange security alerts and other data, where the credibility of peers in a CIDN is crucial.

Establishing and maintaining mutual trust between the IDS nodes is a prerequisite for maintaining a high-level of security, as nodes that turn malicious may degrade the overall security provided by a CIDN. To achieve this high-level of security, continuous monitoring of nodes' behavior is necessary, together with the implementation of a trust model for mutual evaluation, based on previous behavior \cite{fung08}. Apart from the credibility of the CIDNs' nodes, the trustworthiness of external hosts (IP sources) may also be measured. In this way, incoming traffic to the network can be pre-filtered through packet filtering mechanisms and large-scale attacks, like DDoS attacks, can be more efficiently mitigated \cite{meng17}. Towards that direction, blockchain technology, which has already found several applications in the security domain \cite{kiayias16} could be combined with a CIDN in order to achieve trusted distributed coordination needed among its IDS peers \cite{alexopoulos17}, \cite{meng18}. 

In this paper, we propose an novel architecture for a distributed CIDN that includes mechanisms for evaluating the credibility of a CIDN's nodes and realizing trust-based packet filtering of incoming traffic by external IP sources, depending on the trustworthiness of the latter. These mechanisms are supported by blockchain technology, so as to ensure transparency and accountability, whilst improving robustness against insider threats, as the integrity of information shared among the IDS nodes is guaranteed. Our contributions are summarized as follows.
\begin{itemize}
\setlength\itemsep{2pt}
\item A new trust management scheme is proposed that can be used in a CIDN to evaluate the credibility of IDS peers and the trustworthiness of external IP sources. The modeling of trust allows weighting differently the recently observed behavior, as in \cite{ntemos18}, to adjust trust model's sensitivity to behavioral variations.

\item A blockchain solution is designed for storing the trust scores disseminated by the CIDN nodes along with evidence justifying these scores to enhance the overall security and identify misbehaved IDS nodes. 

\item A novel consensus protocol is proposed that combines proof-of-work and proof-of-stake protocols, based on \cite{duong16}, \cite{Duong2018}, \cite{nakamoto08}, and \cite{nxt14}, to facilitate the secure maintenance of the blockchain by the CIDN nodes.
\end{itemize}
The rest of the paper is organized as follows. In Section \ref{sec.background} we provide the background on intrusion detection systems, collaborative intrusion detection and blockchain technology. Section \ref{sec.proposal} formalizes the proposed architecture for a trust-based and blockchain-enhanced CIDN. Section \ref{sec.consensus} presents in more detail our approach on achieving consensus among the CIDN peers on the information stored on the blockchain. Section \ref{Security} presents how our model reacts in an adversarial environment and Section \ref{sec.conclusions} summarizes our contributions and outlines future work directions.

%----------------------------------------------------------%
%                      Section Change                      %
%----------------------------------------------------------%
\section{Background and related work}
\label{sec.background}

This section provides background information on collaborative intrusion detection, trust management schemes and their applications in CIDNs, as well as, on blockchain protocols and consensus mechanisms along with the recent proposals for their use in CIDNs.

%----------------------------------------------------------%
%                      Section Change                      %
%----------------------------------------------------------%
\subsection{Collaborative intrusion detection}
\label{sec.ids.bkg}

Intrusion detection systems are widely used to ensure the security of networks and hosts by collecting and analyzing data for ongoing threats. Detection by an IDS may be either {\em anomaly-based} or {\em signature-based} \cite{garcia09}, \cite{axelsson00}. IDSs may also be classified into {\em host-based} (HIDS), where only one device is monitored, and {\em network-based} (NIDS), where the network traffic is monitored and analyzed \cite{alexopoulos17}. Nevertheless, as stand-alone IDSs are not able to identify large-scale attacks, the use of CIDNs has been proposed \cite{fung11}. A CIDN consists of several monitors, for collecting and sharing security-related data, as well as analysis units for extracting threat intelligence information \cite{wu03}. 

\begin{figure}[t]
\centering
\includegraphics[width=\linewidth]{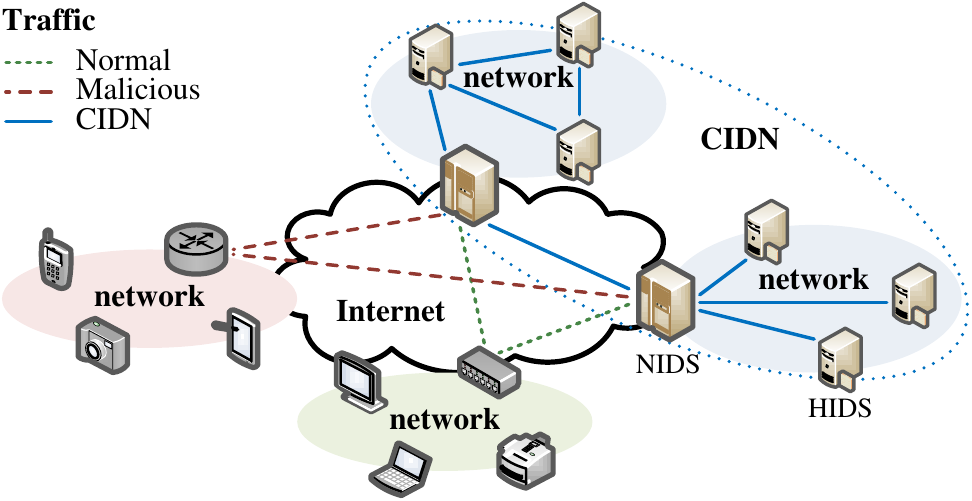}
\caption{Example of a distributed CIDN topology.}
\label{cidn}
\end{figure}

There are three widely adopted architectures concerning the deployment of CIDNs, namely centralized, decentralized and distributed \cite{vasilomanolakis15}. Nodes of a centralized CIDN are only connected to a central unit that is responsible for the analysis of the collected data. If this single unit stops functioning, then the overall protection system collapses; this is the case of a {\em single point of failure} (SPoF) \cite{fung11}. A decentralized CIDN consists of nodes with a topological structure (e.g. hierarchical), so that the analysis units work as filters forwarding correlated data to the higher levels of the network; bottlenecks have also been observed in such architectures. On the other hand, the nodes of a distributed CIDN, as illustrated in Fig. \ref{cidn}, are designed to both collect and analyze data; therefore, all the CIDN nodes have the ability to communicate in a {\em peer-to-peer} (P2P) fashion and achieve significant performance gains towards detecting attacks \cite{locasto05}.

%----------------------------------------------------------%
%                      Section Change                      %
%----------------------------------------------------------%
\subsection{Trust management}
\label{sec.trust.bkg}

Several types of insider attacks are encountered within the CIDN framework, the predominant of which is that of malicious nodes that on purpose share fake data with their peers to significantly deteriorate the performance of the CIDN, and thus a network's security \cite{fung11}. For the mitigation of such attacks within a distributed CIDN, it is suggested that information from only trusted peers should be taken into account. Thus, different schemes of trust management among CIDN nodes have been developed. Duma et. al \cite{duma06} suggested that the IDS nodes continuously monitor the behavior of their CIDN peers and evaluate the quality of the security-related information that they share to estimate their \emph{credibility}. Then, any data contributed by a CIDN node is taken into consideration depending on the node's calculated credibility.

Apart from measuring the credibility of a CIDN's nodes, calculating and maintaining the \emph{trustworthiness} of external IP sources has also been proposed \cite{meng17}. By filtering their incoming packets according to a collaborative trust-based scheme, large-scale DDoS attacks can be mitigated. The packet filtering mechanism is based on maintaining a table of blacklisted (i.e. untrusted) IP sources whose packets are immediately dropped, without further inspection and analysis, thus reducing the workload of detection units. The challenge in the case of distributed CIDNs is that there is no central trusted authority to support the establishment of trusted coordination between the CIDN peers \cite{meng18}. In the sequel, we address this challenge by proposing a solution that relies on the blockchain technology, in order to secure the information on the trustworthiness of external hosts shared by CIDN peers.

%----------------------------------------------------------%
%                      Section Change                      %
%----------------------------------------------------------%
\subsection{Distributed ledgers}
\label{sec.blockchain.bkg}

The blockchain was introduced with the Bitcoin as part of the solution that tackles in a distributed fashion the double-spending problem in a trustless P2P network \cite{nakamoto08}. To achieve this, the solution relies on cryptographic schemes ensuring the immutability of the data records that are stored on the distributed ledger, referred to as {\em transactions} $\Tx$. Moreover, a security through transparency approach is taken, based on which all nodes' transactions are publicly announced, hence allowing anyone to verify their validity. A hash function, e.g. \textsf{SHA}--$256$, is the core cryptographic primitive upon which the security of the whole blockchain construction relies. 

Hash functions are involved in digitally signing $\Tx$ with the {\em private key} of the originator (a CIDN peer in our context); therefore, its authenticity is verified by using the associated {\em public key} that is also included in the blockchain. A number of new $\Tx$ is packed into a block, containing links to past $\Tx$ appearing on the blockchain, and is subsequently appended to the structure. In addition to the above, a block commonly includes the hash of the previous block, a timestamp proving that the data to store exist at a particular time instant and are authentic, as well as, a $\Nonce$ value that is used according to the consensus mechanism.

The mutual agreement on the validity of a newly created block is performed according to a {\em consensus} protocol. This also ensures that tampering or removal of the blocks on the ledger is impossible, thus making the whole data structure immutable. There is a plethora of different mechanisms that have been proposed for achieving consensus; the protocols being relevant to this work are proof-of-work \cite{nakamoto08} and proof-of-stake \cite{nxt14}, \cite{li17}.

Once transactions are validated and inserted into a block, it is exponentially hard for an adversary to alter the contents of that block after it is being appended to the blockchain. In fact, the success probability of such an attack exponentially decreases with the number of blocks that have to be altered by an adversary targeting at a specific block of some depth in the chain \cite{eyal14}. The maintenance of the ledger, i.e. the validation of new transactions, their aggregation into blocks, and the chaining with the structure, is carried out by a class of network nodes (i.e.\ a subset of the CIDN in our context) that depends on the type of the blockchain. Distributed ledgers can be classified as \emph{permissionless} or \emph{permissioned} depending on whether the block generation process is open to all network nodes.

%----------------------------------------------------------%
%                      Section Change                      %
%----------------------------------------------------------%
\section{Proposed architecture}
\label{sec.proposal}

In this section we present the proposed distributed CIDN model for realizing a trust-based packet filtering mechanism that relies on the blockchain technology is presented. 

%----------------------------------------------------------%
%                      Section Change                      %
%----------------------------------------------------------%
\subsection{CIDN model}
\label{sec.ids.model}

\begin{figure}[t]
\centering
\includegraphics[width=\linewidth]{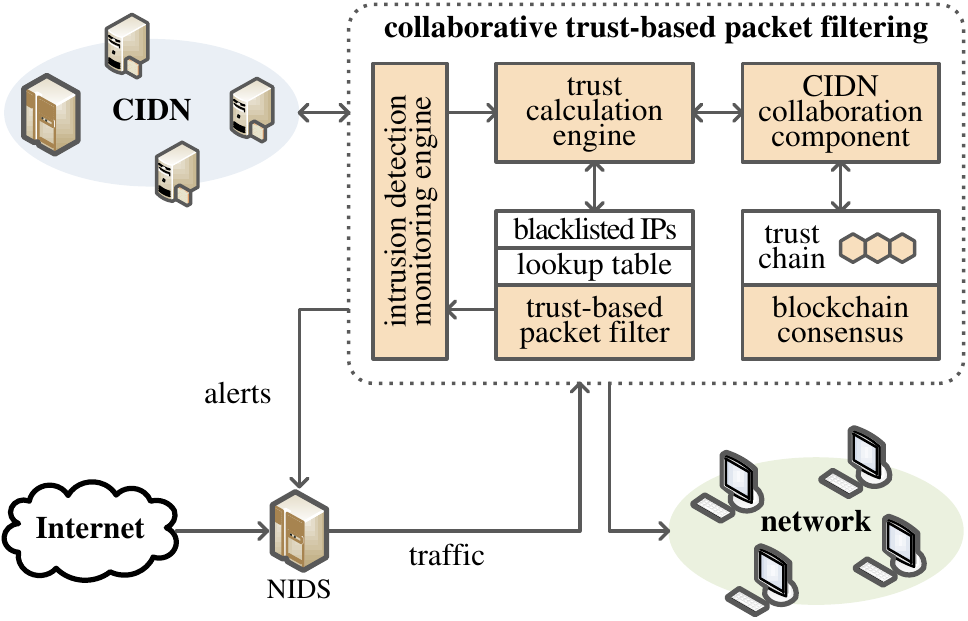}
\caption{High-level view of the collaborative trust-based IDS, enriched with a trust-chain.}
\label{tids}
\end{figure}

Let us assume a CIDN whose members, i.e. the peer IDS nodes, comprise the set $N$. Furthermore, let $M$ be the set of the external network hosts (in the form of IP addresses) that are collectively monitored by the CIDN. We use $M_i \subseteq M$ to denote the subset of IPs monitored by the IDS node $i \in N$, and we define $N_i = N \setminus\{i\}$. The high-level architecture is illustrated in Fig. \ref{tids}, where the primary building blocks of the proposed solution are: (a) the intrusion detection monitoring engine; (b) the CIDN collaboration component; (c) the trust calculation engine; (d) the trust-based packet filter; and (e) the blockchain component referred to as trust-chain.

The baseline functionality of an IDS, part of a CIDN, is structured upon the first two components, the intrusion detection monitoring engine and the collaboration component. Similarly to the scheme proposed in \cite{meng17}, the trust calculation engine is used to keep track of the behavior of the various entities involved (members of $N$ and $M$), while the packet filter is the component where incoming data filtering takes place. Part of the latter are also a list of blacklisted IP addresses and a set of signatures, against which the incoming packets are compared to. Incoming packets may be dropped if their source IP address is blacklisted. In addition, the trust calculation engine continuously updates the blacklist by collecting alerts from the detector and side information from the CIDN peers about the trustworthiness of the external host. The binary decision on inclusion or exclusion of the host into the blacklist, thus its trustworthiness, relies on the comparison against a threshold $\zeta \in (0,1)$. Therefore, the IDS node handles only the accepted packets, practically reducing the load of an IDS during operation. The collaboration component communicates with the other peers to transfer security-related data.

The last component, which is a key contribution of this work on the CIDN architecture proposed in \cite{meng17}, is the so-called trust-chain. It is comprised by the specific blockchain structure along with the associated consensus protocol that are described in detail in Sections \ref{sec.tchain} and \ref{sec.consensus} respectively. This structure is where the information shared about the trustworthiness of the external hosts is stored.

%----------------------------------------------------------%
%                      Section Change                      %
%----------------------------------------------------------%
\subsection{Trust engine}
\label{sec.trust.model}

In order to deal with internal and external attacks from misbehaving nodes, two types of trust scores are considered that characterize the {\em credibility} of an IDS peer (member of the CIDN) and the {\em trustworthiness} of an external host. The above notions were also used in \cite{fung11}, \cite{fung08}, but the subsequent formulation of the trust model differs in many aspects. To be more precise, main design choices of \cite{ntemos18} are adopted as the trust model proposed therein was shown to be quite robust in an adversarial environment. The parameters used are
\begin{itemize}
\setlength{\itemsep}{2pt}
\item the forgetting factor $\lambda \in (0,1)$ controlling the weight given to past behavior;

\item the severity level $\phi > 0$ defining the punishment of an IDS node that avoids giving feedback to challenges;

\item the credibility threshold $\theta \in (0,1)$ that determines the IDS nodes whose data on the trust-chain are trusted;

\item the initial trust score $\tau \in [0,1]$ assigned to a new IDS node when entering the CIDN network.
\end{itemize}
The credibility of an IDS node $i \in N$ relies on its responses to challenges (alert priorities) that are sent out periodically following a probability distribution and are indistinguishable from real alerts. The format of the challenges may adhere to the \emph{intrusion detection message exchange format} (IDMEF) standard \cite{debar07}. The responses given to these challenges are used to compute the requesting node's $j$ {\em satisfaction} level $\sat_j(i) \in [0,1]$, which depends on the gap between the actual and the expected responses \cite{fung11}, \cite{fung08}. These values are combined using a forgetting factor to derive an {\em accumulated satisfaction} level $\gamma_j(i) \in [0,1]$ as follows
\begin{equation}
\label{eq1}
\gamma_j(i) = \bigl( 1 - \lambda \bigr) \cdot \sat_j(i)  + \lambda \cdot \gamma_j(i)\,, \quad i\in N_j\,.
\end{equation}
Honest IDS nodes always respond correctly to challenges, if information about the matched item exists, or they respond with $\Unsure$ otherwise. Let us define the variable  $\ans_j(i) \in \{0,1\}$ that equals $1$ if and only if an $\Unsure$ response is given by the IDS node $i$ to $j$. To avoid having malicious IDS nodes abuse the ability to respond with $\Unsure$, instead of forcing to guess the challenge's correct answer (thus leading to a decreased satisfaction level), the quantity $\alpha_j(i) \in [0,1]$ is computed recursively by the expression
\begin{equation}\label{eq3a}
\alpha_j(i) = \bigl( 1 - \lambda \bigr) \cdot \ans_j(i)  + \lambda \cdot \alpha_j(i)\,, \quad i\in N_j
\end{equation}
and accounts for the percentage of the $\Unsure$ responses that IDS node $i$ gives to $j$. Then, according to $j$, the credibility $\crd_j(i) \in [0,1]$ of the IDS node $i$ is computed based on the severity of punishment $\phi$ for providing $\Unsure$ answers
\begin{equation}\label{eq3b}
\crd_j(i) = \bigl( 1- \alpha_j(i) \bigr)^\phi \cdot \bigl( \gamma_j(i) - \tau \bigr) + \tau\,, \quad i\in N_j
\end{equation}
whereas $\crd_j(j) = 1$, for all $j\in N$, by definition. From \eqref{eq3b} we obtain that $\crd_j(i) = \tau$ when the IDS node $i$ constantly responds with $\Unsure$. Assuming that only IDS peers whose credibility exceeds a threshold $\theta$ are taken into account when incorporating the knowledge acquired by the whole CIDN, the relative weight given by $j$ to the IDS node $i\in N$ is
\begin{equation}
w_j(i) =
\begin{cases}
0 \,, & \text{if}\ \crd_j(i) < \theta \\
\crd_j(i)\hspace*{-2pt}\Bigg/\hspace*{-8pt}{\displaystyle\sum_{\substack{l \in N:\\ \crd_j(l) \geq \theta}} \hspace*{-4pt} \crd_j(l)} \,, & \text{otherwise}
\end{cases}
\end{equation}
where we clearly have $\sum_{i \in N} w_j(i) = 1$ amongst the nodes comprising the collaborative intrusion detection network $N$.

In order to determine the trustworthiness of an external host $\ip \in M_j$, the IDS node $j$ monitors the traffic that is received by the particular host and detects whether a data packet is normal or malicious. At regular intervals, every $n$ data packets, the trust calculation engine computes a belief about the type of the next packet; assuming that $k$ out of the $n$ packets were detected to be normal, then the probability that the $(n+1)$th packet is normal equals \cite{meng17}
\begin{equation}\label{eq.measure.trust}
\begin{aligned}
\tr_j(\ip) 
&= \Pr(n + 1 = \text{normal}\ \big|\ k \ \text{normal} \,) \\
&= \frac{1 + k}{2 + n}
\end{aligned}
\end{equation}
when the distribution of observing $k$ normal packets (out of the $n$ packets) is the Binomial distribution. This score is the observation of the IDS node $j$, as measured during the last monitoring interval, and is referred to as the {\em instantaneous} trust score $\tr_j(\ip) \in (0,1)$ of the external host $\ip$. Likewise, the IDS node $j$ may calculate the long-term or {\em accumulated} trust score \zbx{$\tr_j^{\ids}(\ip)$} of the host using the expression
\begin{equation}\label{eq.update.local}
\tr_j^{\ids}(\ip) = \bigl( 1 - \lambda \bigr) \cdot \tr_j(\ip) + \lambda \cdot \tr_j^{\ids}(\ip)
\end{equation}
that incorporates the past knowledge that the IDS has about the particular host. However, in order to take advantage of the collective knowledge that the whole CIDN has about $\ip$, the IDS node $j$ utilizes the individual trust scores that have been computed locally by other credible peers of the CIDN. This process yields the {\em combined} trust score \zbx{$\tr_j^{\cidn}(\ip)$} whose computation is based on the weighted combination
\begin{equation}\label{eq.compute.combined}
\tr_j^{\cidn}(\ip) = \sum_{i \in N}   w_j(i) \cdot  \tr_i^{\ids}(\ip)
\end{equation}
summarizing the trustworthiness of host $\ip$ using the CIDN's knowledge, as aggregated by the IDS node $j$. In the final step of the above process, the IDS updates its internal value \zbx{$\tr_j^{\ids}(\ip)$} with the one computed in \eqref{eq.compute.combined}. The host's IP is added to the blacklist if \zbx{$\tr_j^{\ids}(\ip) \leq \zeta$}, and excluded otherwise.

The above sequence of steps is also illustrated in Alg. \ref{alg.trust.distr}; this is a typical {\em adopt-then-combine} scenario for performing in-network processing in diffusion networks, an approach that has proven to be resilient in adversarial environments \cite{ntemos18}. Two phases have been realized in Alg. \ref{alg.trust.distr}: during the first phase, the IDS nodes augment the local knowledge about an external host and disseminate it in the CIDN, whereas in the second phase, the IDS nodes aggregate the updated knowledge (in the form of a trust score) received from their peers. In the sequel, this algorithm is extended in order to allow a secure realization of the information sharing via the trust-chain.

\begin{algorithm}[!t]
%\floatname{algorithm}{Alg.}
\caption{Distributed computation of IPs' trust in a CIDN in a typical adaptive diffusion scenario.}\label{alg.trust.distr}
\begin{algorithmic}[1]
\Require {CIDN nodes $N $, list of IPs $M$}
\Initial {\zbx{$\tr_j^\text{ids}(\ip) \gets \tau$}}
\Comment {$\forall \, j \in N,\, \ip \in M_j$}
\vspace*{4pt}

\For {$t \gets 1,2,\ldots$}

\ForAll {$j \in N,\, \ip \in M_j$}
\State {measure \zbx{$\tr_j(\ip)$}}
\Comment {from \eqref{eq.measure.trust}}
\State {update accumulated \zbx{$\tr_j^\text{ids}(\ip)$}}
\Comment {from \eqref{eq.update.local}}
\State {send \zbx{$\tr_j^\text{ids}(\ip)$}}
\Comment {to all $i \in N$}
\EndFor

\vspace*{4pt}

\ForAll {$j \in N,\, \ip \in M_j$}
\State {receive \zbx{$\tr_i^\text{ids}(\ip)$}}
\Comment {from all $i \in N$}
\State {compute combined \zbx{$\tr_j^\text{cidn}(\ip)$}}
\Comment {from \eqref{eq.compute.combined}}
\State {\zbx{$\tr_j^\text{ids}(\ip) \gets \tr_j^\text{cidn}(\ip)$}}
\EndFor

\EndFor

\vspace*{4pt}
\Ensure {\zbx{$\tr_j^\text{ids}(\ip)$}}
\Comment {$\forall \, j \in N,\, \ip \in M_j$}
\end{algorithmic}
\end{algorithm}

%----------------------------------------------------------%
%                      Section Change                      %
%----------------------------------------------------------%
\subsection{Trust-chain structure}
\label{sec.tchain}

In order to provide a more accountable trust management framework, the IDS node $j$ retains some evidence $\ev_j(\ip)$ after measuring the trustworthiness of an external host $\ip$ to justify the scoring (e.g. alerts having been disseminated to the CIDN during the previous monitoring interval). Hence, the IDS node $j\in N$ maintains the following lists 
\begin{equation}\label{evidencelist}
\begin{aligned}
C_j &= \bigl\{ \crd_j(i):         i \in N_j \bigr\} \\
L_j &= \bigl\{ \tr_j^{\ids}(\ip): \ip \in M_j \bigr\} \\
E_j &= \bigl\{ \ev_j       (\ip): \ip \in M_j \bigr\} 
\end{aligned}
\end{equation}
in addition to the IDS nodes in $N_j$ and external hosts in $M_j$ that are monitored. The transaction $\Tx_j$ is disseminated to all CIDN members, when differences on the credibility of the IDS nodes or the trust-scores of the IP hosts occur, ({\em see} step \ref{alg.step.broadcast} of Alg. \ref{alg.trust.bchain}) and has the following structure
\begin{equation*}
\Tx_j = \id_{\Tx} || \id_{\ids} || \bigl\{ N_j || C_j \bigr\} || \bigl\{ M_j || L_j || E_j \bigl\} || \sig_{\Tx} \,.
\end{equation*}
Each transaction $\Tx_j$ is given a unique identifier $\id_{\Tx}$ and apart from the lists, which constitute the transactions' payload, information is embedded about the identity $\id_{\ids}$ of the IDS and the signature $\sig_{\Tx}$ that is computed with the IDS node's private key. All the transactions in the CIDN during the last monitoring period are denoted as
\begin{equation}\label{tx}
\Tx = \{\Tx_j: j \in N\}
\end{equation}
gathering the information that is disseminated to the CIDN. A number of IDSs having been found to be credible nodes, attempt to generate the new block $B$ that will be appended into the trust-chain and then validated by all the CIDN. This process is detailed in Section \ref{sec.consensus} and corresponds to step \ref{alg.step.manage} of Alg. \ref{alg.trust.bchain}. The block $B$, which is comprised of a header and a payload (i.e.\ all the transactions $\Tx$ defined in \eqref{tx})
\begin{equation}\label{blck}
B = \hdr_B || \Tx
\end{equation}
is signed with the leader's private key. The block's header $\hdr_B$ contains information on the block's unique identifier $\id_B$, the leader's identity $\id_{\ids}$, a time-stamp that verifies the block's generation time $\GenTime$, the hash  $\PrvHash$ of the last block on the trust-chain, a counter $\ctr\leq q$ (where $q$ is the maximum number of attempts to generate a block) and a target value $V_{\ids}$. Thus, the header is structured as
 \begin{equation}
 \label{blck.hdr}
\hdr_B = \id_B || \id_{\ids} || \GenTime || \PrvHash || \ctr || V_{\ids}
\end{equation} 
where $V_{\ids}$, along with other information, allows members of the CIDN to validate the credibility of the IDS node acting as a leader and generating the block $B$. The trust-chain secure process of sharing information on credibility, trustworthiness and the associated evidence is illustrated in Alg. \ref{alg.trust.bchain}.

\begin{algorithm}[!t]
%\floatname{algorithm}{Alg.}
\caption{Distributed computation of IPs' trust in a CIDN with the use of trust-chain.}\label{alg.trust.bchain}
\begin{algorithmic}[1]
\Require {CIDN nodes $N$, list of IPs $M$}
\Initial {\zbx{$\tr_j^\text{ids}(\ip) \gets \tau$}}
\Comment {$\forall \, j \in N,\, \ip \in M$}
\vspace*{4pt}

\For {$t \gets 1,2,\ldots$}

\ForAll {$j \in N$}
\ForAll {$\ip \in M_j$}
\State {measure \zbx{$\tr_j(\ip)$}}
\Comment {from \eqref{eq.measure.trust}}
\State {update accumulated \zbx{$\tr_j^\text{ids}(\ip)$}}
\Comment {from \eqref{eq.update.local}}
\EndFor
\State {build $\Tx_j$ from \zbx{$C_j, L_j, E_j$}}
\State {broadcast $\Tx_j$} \label{alg.step.broadcast}
\Comment {to all $i \in N$}
\EndFor

\vspace*{4pt}
\State {generate block $B$}
\Comment {from \eqref{blck}}
\State {manage trust-chain}
\Comment {consensus protocol}
\label{alg.step.manage}
\vspace*{4pt}

\ForAll {$j \in N$}
\State {extract $\Tx_i$ from $B$}
\Comment {from all $i \in N$}
\State {read \zbx{$C_i, L_i, E_i$} from $\Tx_i$}
\ForAll {$\ip \in M_j$}
\State {compute combined \zbx{$\tr_j^\text{cidn}(\ip)$}}
\Comment {from \eqref{eq.compute.combined}}
\State {\zbx{$\tr_j^\text{ids}(\ip) \gets \tr_j^\text{cidn}(\ip)$}}
\EndFor
\EndFor

\EndFor

\vspace*{4pt}
\Ensure {\zbx{$\tr_j^\text{ids}(\ip)$}}
\Comment {$\forall \, j \in N,\, \ip \in M$}
\end{algorithmic}
\end{algorithm}

%----------------------------------------------------------%
%                      Section Change                      %
%----------------------------------------------------------%
\section{Trust-chain's consensus}
\label{sec.consensus}

In this section we present the details of electing an IDS node that is credible enough for generating the next block in the trust-chain. We propose a solution combining the PoW and PoS protocols to achieve consensus \cite{bentov14}, \cite{duong16}, where the PoS protocol extends that of Nxt \cite{nxt14}. High-level functions that realize the core functionality of the proposed solution are presented below.

\setlength\partopsep{4pt}
\begin{description}[\setleftmargin{0pt}\setlabelphantom{xx}]
\setlength\itemsep{4pt}
\item [$\mathsf{CheckEligibility} \big(\id_j, D, \overline{\crd}(j), \PrvHash, \Tx \big)$] checks if an IDS node $j$ is eligible for generating the next block; it takes as input the identity of the IDS, a system parameter $D$, the average credibility placed on $j$ from the CIDN, the hash value of the previous block and the payload; it computes a hash value $\mathcal{G}_j$ that is used next and outputs a Boolean value ($\True$ or $\False$).

\item [$\mathsf{GenerateBlock} \big( \mathcal{G}_j, \GenTime, \ctr, D', \Stake_j, \Time_j \big)$] adds a new block $B$ in the trust-chain; the input to this function are the hash value $\mathcal{G}_j$ provided by the $\mathsf{CheckEligibility}$ function, a time-stamp verifying the block's generation time, a counter $\ctr$, a target value $D'$ (different from $D$), the $j$th node's stake $\Stake_j$, and the time elapsed $\Time_j$ since the last block generated by $j$. It outputs a Boolean value ($\True$ or $\False$).

\item [$\mathsf{ValidateBlock} \big( B, D, D' \big)$] with input the block $B$ and the target difficulties $D,D'$, validates block $B$ and returns $\True$ if and only if the output of the $\mathsf{CheckEligibility}$ and the $\mathsf{GenerateBlock}$ functions are both $\True$. 

\item [$\mathsf{Resolve} \big(\fork_1, \ldots, \fork_z \big)$] returns the unique fork to work on, assuming that a number $z$ of forks has been detected. This is the case where multiple IDS nodes satisfy the conditions of the election process for generating the next block. 
\end{description}

\noindent
According to the philosophy of PoS (resp. PoW) protocols, the node with the highest stake (resp. computational power) is more likely to generate the next block. The combination of PoW and PoS protocols in trust-chain leads to a hybrid mining-election method for achieving consensus, where the likelihood of a credible IDS $j$ being elected as the leader increases with both its computational power and stake. This explains the inclusion of a counter $\ctr$ into the block $B$ as given in \eqref{blck.hdr}. The advantage of the combination is that it prevents situations in which a credible IDS node with large stake is in position to ceaselessly generate all the blocks. 

In the context of collaborative intrusion detection that is based on credibility placed between IDS nodes, each peer can take advantage of its behavior  and its contribution to the CIDN. The information in the list $C_j$, maintained by every IDS node $j$ on the other peers' credibility, is disseminated to the CIDN through the trust chain and the average credibility score for $j\in N$ is then computed as follows
\begin{equation}
\label{eq.average}
\begin{aligned}
\overline{\crd}(j)= \frac{1}{|N|} \left( 1 + \sum_{l \in N_j} \crd_l(j) \right)
\end{aligned}
\end{equation}
at each time interval. It is then used to decide whether $j$ is considered to be credible enough by the whole CIDN to be elected as a leader for the next block generation. Thus, IDS node $j$ first computes the output $\mathcal{G}_j$ of a hash function $\mathsf{G}(\cdot)$ and then checks if the {\em credibility condition} is satisfied
\begin{equation} 
\label{eq.credibility}
\begin{aligned}
\mathcal{G}_j = \mathsf{G}\bigl(\, \id_j, \PrvHash, \Tx \,\bigr) < D \cdot \overline{\crd}(j)
\end{aligned}
\end{equation}
based on a  target $D$ and the  average credibility score placed on node $j$ by the CIDN. Note that it is important to adjust $D$ as a loose difficulty target so as to allow many IDS nodes with high average credibility to participate in the following process and satisfy the {\em mining-election condition} 
\begin{equation} \label{eq.ineq2}
\begin{aligned}
\mathsf{Prefix}\bigl(\, \mathsf{H}\bigl(\, \mathcal{G}_j, \GenTime, \ctr \,\bigr), r \,\bigr) < V_j \,.
\end{aligned}
\end{equation}
To achieve consensus, the IDS nodes brute-force \eqref{eq.ineq2} using the available computational resources by continuously trying different values for $\ctr$ and comparing the first $r$ bits of the hash resulting from $\mathsf{H}(\cdot)$ with $V_j$, where
\begin{equation}
\label{eq.stake}
\begin{aligned}
V_j= D' \cdot \Stake_j \cdot \Time_j \,.
\end{aligned}
\end{equation}
Then, an IDS node $j$ generates the next block if and only if both \eqref{eq.credibility}, \eqref{eq.ineq2} hold. In addition to the time elapsed since the IDS node $j$ has generated the last block, its winning chance is linked to its ability in detecting the trustworthiness of the external hosts monitored so as to improve the accuracy of the trust-based packet filter employed by each CIDN node. This is captured by the uncertainty that the IDS $j$ has when assigning a trust score \zbx{$x = \tr_j^{\ids}(\ip)$}. Thus, we let
\begin{equation}
\Stake_j = \sum_{ x \in L_j } \bigl( 1 - H_2(x) \bigr)
\end{equation}
where the index $x$ runs through all the trust scores in the list \zbx{$L_j$} while $H_2(x) = - x \log_2(x) - (1-x) \log_2(1-x)$ is the binary entropy function. By definition, the closer to $1/2$ the trust scores in $L_j$ are, the less useful they will be in arguing about the trustworthiness of a host.

Since a combined PoW and PoS consensus protocol is proposed, a new block is more accurate if it is generated by a credible node. Furthermore, the efficiency of the leader to monitor and calculate the trust values of the external hosts  is more important than its computational power. Therefore, the wining chain, when forks occur, is the one that possesses the highest accumulated stake by the most credible nodes and thus avoid insider attacks based on computational power.

%----------------------------------------------------------%
%                      Section Change                      %
%----------------------------------------------------------%
\section{Discussion on trust-chain's security }
\label{Security}

A trust management scheme combined with the properties of the blockchain can adequately improve the collaboration among IDS nodes. Nevertheless, malicious behavior, can possibly degrade the efficiency of the entire system. In this section, we describe a number of attacks and present how the proposed system provides proper defenses against them. Thwarting insider attacks is a challenging task in collaborative security mechanisms; they distribute false information to manipulate the outcome of a system's or peer's aggregation function. Attackers may penetrate a CIDN while acting as credible parties, to perform some security-related tasks, and thus disturb and obstruct the normal decision-making of the whole system. These attacks may be categorized according to three different criteria.

\begin{itemize}
\setlength{\itemsep}{4pt}
\item {\em Type of attack:} the attacks can be subdivided into (a) those targeting at the identities of the CIDN nodes, (b) those being related to the data exchanged amongst the nodes, and (c) the attacks that target the routing of data among the nodes. 

\item {\em Attacker's behavior:} in case of multiple attackers, they can either act independently, i.e. the malicious actions taken serve each attacker's own purposes, or they may collaborate with each other, leading to collusive attacks.

\item {\em Attacker's intelligence:} in the simplest scenario and in most works in the literature, the attackers' behavior is static as they just repeat a particular type of malicious action. At the extreme end, attackers can be intelligent and they change their tactics strategically to avoid being detected or to maximize the attack's impact. Finally, the attackers may behave irrationally, thus preventing their behavior being predicted.
\end{itemize}

\noindent
The proposed solution relies on blockchain technology to build the so-called trust-chain, which aims at protecting the integrity of the information shared among the CIDN peers, enhance their accountability, and secure their collaboration by thwarting insider attacks. The proposed consensus protocol, which is a combination of the PoS and PoW protocols, enables collaborative IDS nodes to maintain a reliable and tampered-resistant trust-chain. The prominent attacks in our setup include Sybil, betrayal and collusion attacks.

In a Sybil attack, malicious IDS nodes create several fake identities to gain larger influence on alert dissemination and aggregation and block the propagation of certain messages \cite{fung11}, \cite{eyal14}. In our system, the IDS members are authenticated and newcomers (possibly fake nodes) have to contribute to the CIDN to gain credibility before given the opportunity to generate the next block of the trust-chain. Assuming that all IDS nodes have the same hashing power, then the probability that one is elected to generate the next block is proportional to its stake as well as the average credibility placed on it by the CIDN.

During a betrayal attack, a (usually highly) credible node gets compromised and subsequently turns malicious. Then, it can either act independently or in collaboration with other malicious IDS nodes \cite{2006-duma}. To defend this attack, a forgetting factor and a severity punishment are implemented into our scheme so that the credibility of a malicious node drops fast enough after few abnormal actions. In this case, the average credibility placed on a specific peer is reduced, abbreviating its opportunity to create the next block. In addition, a counter $\ctr$ has been included in the block generation process so as to enable the participation of only credible IDS nodes, while preventing nodes with large stake from generating sequences of consecutive blocks in the trust-chain.

In a collusion attack a set of dishonest IDS nodes might cooperate to tamper the trust-chain. This can be done either by intentionally broadcasting malicious messages (i.e.\ alerts, trust-scores about the IP hosts and the IDS nodes and false evidence) throughout the CIDN network, or by using their power (hashing or stake) to generate an adversarial block. In the first case, each IDS depends on its own experience to unmask the adversaries using challenges (test messages, priority alerts), that are sent in a random way and are difficult to be distinguished by actual alerts. On the other hand, when a group of malicious nodes works together may have a great impact on the security of the network and the blockchain. Adversarial nodes can use their resources and become the only block generators in the network thus forcing the honest nodes to work for nothing \cite{eyal14}. However in our hybrid PoW and PoS protocol, the adversary needs to control not only a great portion of the hashing power, but also a large portion of the total network's stake and the majority of credible IDS nodes. Each valid block has to be created by an authenticated and credible IDS node based on different parameters (stake, hashing power, as well as time elapsed since the last block creation). In the worst case scenario, where an adversary is able to create a malicious block and fork the trust-chain, the $\mathsf{Resolve}$ function is utilized and returns the fork created by the most credible nodes with the highest accumulated stake. Therefore, a collusion attack is highly unlikely to occur.

Since the security of the proposed blockchain mechanism is of utmost importance, a plethora of fundamental properties need to hold, such as persistence, liveness, chain quality, and the common prefix property \cite{kiayias16}, \cite{kiayias15}; their formal analysis is outside the scope of the present work and constitutes part of ongoing research. If all true,  the ability of the adversaries to alter trust-chain's evidentiary data would be considerably limited.

%----------------------------------------------------------%
%                      Section Change                      %
%----------------------------------------------------------%
\section{Conclusions}
\label{sec.conclusions}

In this paper, a distributed trust management framework for CIDNs was proposed. More precisely, each IDS shares trust-related information about IDS nodes and external hosts with other CIDN members by using an adopt-then-combine approach; this information is securely aggregated according to the source IDS's credibility, which is computed based on the responses given to challenges. The security-related data that have to be exchanged between members of the CIDN is stored on a blockchain, referred to as trust-chain, to avoid tampering from malicious nodes. A combined PoW and PoS protocol was proposed, according to which a credible IDS node with higher computational power and larger stake has an increased probability of being elected for the generation of the next block.
 
Ongoing work focuses on the theoretical aspects of security, namely to study a series of attacks having been reported in both domains (trust management and blockchain), so as to fully understand the impact of various parameter choices on the proposed solution's security and the dynamics governing the trust score evolution. Simulations on the proposed system and privacy issues are open problems that will be presented in detail in a forthcoming work.

%----------------------------------------------------------%
%                      Section Change                      %
%----------------------------------------------------------%
\bibliographystyle{IEEEtran}
\bibliography{IEEEabrv,Paper}

\end{document}